\documentclass[journal]{IEEEtran}
\usepackage{cite,amsmath,amssymb,pifont}
\usepackage{color}
\usepackage{graphicx}
\usepackage{pstool}
\usepackage{array,booktabs,multirow}

\usepackage{dblfloatfix} 

\usepackage[subrefformat=parens,labelformat=parens]{subfig}

\newcommand{\ve}[1]{\mathbf{#1}}
\newcommand{\te}[1]{\overline{\overline{#1}}}

\newcolumntype{N}{@{}m{0pt}@{}}

\newcounter{tempEquationCounter}
\newcounter{thisEquationNumber}

\begin{document}

\title{Fundamental Properties and Classification of \\ Polarization Converting Bianisotropic Metasurfaces}

\author{Karim~Achouri and Olivier J. F. Martin}



\maketitle

\begin{abstract}
We provide a detailed discussion on the electromagnetic modeling and classification of polarization converting bianisotropic metasurfaces. To do so, we first present a general approach to compute the scattering response of such metasurfaces, which relies on a generalized sheet transition conditions based susceptibility model. Then, we review how the fundamental properties of reciprocity, energy conservation, rotation invariance and matching may be expressed in terms of metasurface susceptibilities and scattering parameters, and show how these properties may affect and limit the polarization effects of metasurfaces. Finally, we connect together the metasurface susceptibility model to the structural symmetries of scattering particles and their associated polarization effects. This work thus provides a detailed understanding of the polarization conversion properties of metasurfaces and may prove to be of particular interest for their practical implementation.
\end{abstract}	


\IEEEpeerreviewmaketitle


\section{Introduction}

Bianisotropic metasurfaces are electrically thin periodic arrays of scattering particles engineered to provide compact, efficient and advanced electromagnetic wave control capabilities~\cite{achouri2018design,chenHuygensMetasurfacesMicrowaves2018,asadchy2018bianisotropic}. Some of their most remarkable features are enabled by bianisotropy, which is a property requiring coupling between an electric (magnetic) excitation and a magnetic (electric) induced response, that is instrumental to the design of certain metasurface transformations such as polarization rotation via chirality~\cite{Sihvola1994,serdkov2001electromagnetics}, and perfect refraction via asymmetric matching~\cite{Asadchy2016a,8259235}. 

In the context of this work, we are mostly interested in the polarization converting capabilities of metasurfaces, which, in recent years, have led to a plethora of metasurface concepts and applications, such as, for instance, collimating lense for circularly polarized waves~\cite{phillion2011lenses}, polarization dependent excitation of surface waves~\cite{lin2013polarization}, chiral based electromagnetic absorber~\cite{li2014ultra}, chiral polarization control~\cite{kim2016highly}, full-stokes imaging polarimetry\cite{arbabi2018full} and tunable polarization rotation~\cite{wu2019tunable}. Meanwhile, several metasurface synthesis techniques, specifically aimed at polarization control, have been developed to provide guidelines and optimized implementation procedures for the practical realization of efficient metasurfaces~\cite{6477089,PhysRevApplied.2.044011,selvanayagam2014polarization,selvanayagam2015design,kamali2018review}. These techniques describe how polarization conversion in metasurfaces may be modeled using effective material parameters such as polarizabilities or impedances. They also provide design strategies to realize these metasurfaces using specific types of metallic or dielectric scattering particles. On the other hand, several studies have investigated how the shape of the scattering particles, and their related structural symmetries, may affect the scattering response and associated polarization effects of metasurfaces~\cite{menzel_advanced_2010,kenanakis2012flexible,kruk2020tailoring}. However, a general discussion connecting together the structural symmetries of scattering particles and their scattering effects to a bianisotropic electromagnetic model of metasurfaces is still missing in the literature.

This work thus aims at filling this gap by providing a detailed discussion on the fundamental properties of polarization converting metasurfaces. For this purpose, we extend the metasurface modeling framework based on bianisotropic susceptibility tensors developed in~\cite{achouri2014general,angularAchouri2020,achouri2020electromagnetic} by investigating how the fundamental properties of reciprocity, energy conservation, rotation invariance and matching, affect the polarization converting capabilities of metasurfaces, and provide general relationships between the susceptibilities, the structural symmetries of scattering particles and their polarization effects.

This paper is organized as follows. Section~\ref{sec:GSTC} reviews the general modeling procedure for bianisotropic metasurfaces and shows how the susceptibilities of a spatially uniform metasurface may be related to its scattering parameters. Section~\ref{sec:cond} presents the fundamental properties of reciprocity, energy conservation, rotation invariance and matching, and provides the associated conditions in terms of susceptibilities and scattering parameters. Section~\ref{sec:combined} investigates how the aforementioned conditions may be combined with each other and, for each possible resulting case, derives a set of specific related conditions, whose application is illustrated with a metasurface synthesis example pertaining to polarization conversion. Then, Sec.~\ref{sec:sym} describes how the structural symmetries of scattering particles affect the polarization effects of metasurfaces. Finally, Sec.~\ref{sec:concl} concludes the discussion.

\section{GSTC Modeling of Metasurfaces}
\label{sec:GSTC}

Consider a metasurface lying in the $xy$-plane at $z=0$. The interactions of the metasurface with the fields of incident and scattered waves may be modeled using the zero-thickness generalized sheet transition conditions (GSTCs)~\cite{Idemen1973,kuester2003av}, as\footnote{The time dependence $e^{j\omega t}$ is assumed throughout.}
\begin{subequations}\label{Eq:FD_GSTC}
	\begin{align}
	\hat{\ve{z}}\times\Delta\ve{H}&=j\omega\ve{P_{\|}}-\hat{\ve{z}}\times\nabla_{\|}M_{z},\label{eq:CurlH}\\
	\hat{\ve{z}}\times\Delta\ve{E}&=-j\omega\mu_0\ve{M_{\|}}-\hat{\ve{z}}\times\nabla_{\|}(P_{z}/\epsilon_0),\label{eq:CurlE}
	\end{align}
\end{subequations}
where $\Delta\ve{H}$ and $\Delta\ve{E}$ are the differences of the magnetic and electric fields between both sides of the metasurface, $\ve{P}$ and  $\ve{M}$ are electric and magnetic surface polarization densities induced on the metasurface and $\parallel$ refers to components tangential to the metasurface plane. 

For a bianisotropic metasurface, the surface polarization densities in~\eqref{Eq:FD_GSTC} may be expressed in terms of surface susceptibility tensors as~\cite{Kong2008,kuester2003av}
\begin{subequations}
	\label{eq:bianiconstiPM}
	\begin{align}
	\ve{P} &= \epsilon_0\te{\chi}_\text{ee}\cdot\ve{E}_\text{av} + \frac{1}{c_0}\te{\chi}_\text{em}\cdot\ve{H}_\text{av},\\
	\ve{M} &= \te{\chi}_\text{mm}\cdot\ve{H}_\text{av} + \frac{1}{\eta_0}\te{\chi}_\text{me}\cdot\ve{E}_\text{av},
	\end{align}
\end{subequations}
where $\eta_0$ and $c_0$ are the impedance and speed of light in vacuum, $\ve{E}_\text{av}$ and $\ve{H}_\text{av}$ are the average electric and magnetic fields at the metasurface, and $\te{\chi}_\text{ee}$, $\te{\chi}_\text{mm}$, $\te{\chi}_\text{me}$ and $\te{\chi}_\text{em}$ are respectively the electric, magnetic, magnetic-to-electric and electric-to-magnetic metasurface susceptibility tensors, which are here $3\times 3$ matrices.

The most common applications of the GSTCs~\eqref{Eq:FD_GSTC}, along with the constitutive relations~\eqref{eq:bianiconstiPM}, are the synthesis of metasurfaces, which consists in expressing the metasurface susceptibilities in terms of specified incident, reflected and transmitted fields, as well as the analysis of metasurfaces, which consists in computing the fields scattered by a metasurface with known susceptibilities~\cite{achouri2014general,achouri2018design,vahabzadehComputationalAnalysisMetasurfaces2018,achouri2020electromagnetic}.  

In this work, we are rather interested in investigating several fundamental properties of metasurfaces and revealing how they pertain to polarization conversion. For this purpose, we next restrict our attention to \textit{uniform} metasurfaces, i.e., metasurfaces that do not change the direction of wave propagation\footnote{We consider that the metasurface is made of a subwavelength periodic lattice of scattering particles so that all diffraction orders are suppressed except for the 0$^\text{th}$-order ones in reflection and/or transmission.}. Moreover, we also limit our developments to the case of normally impinging plane waves as a source of excitation. Under these conditions of uniformity and normal plane wave incidence, the spatial derivatives in~\eqref{Eq:FD_GSTC} vanish and the presence of normal polarizations may be ignored since they do not contribute to metasurface scattering, as discussed in~\cite{achouri2020electromagnetic,angularAchouri2020}.

Substituting~\eqref{eq:bianiconstiPM} into~\eqref{Eq:FD_GSTC} and removing all spatial derivatives, thus yields
\begin{subequations}
	\label{eq:InvProb}
	\begin{align}
	\ve{\hat{z}}\times\Delta\ve{H}
	&=j\omega\epsilon_0\te{\chi}_\text{ee}\cdot\ve{E}_\text{av}+jk_0\te{\chi}_\text{em}\cdot\ve{H}_\text{av},\label{eq:diffH}\\
	\ve{\hat{z}}\times\Delta\ve{E}
	&=-j\omega\mu_0 \te{\chi}_\text{mm}\cdot\ve{H}_\text{av}-jk_0\te{\chi}_\text{me}\cdot\ve{E}_\text{av},\label{eq:diffE}
	\end{align}
\end{subequations}
where $k_0$ is the wavenumber in vacuum. Since we are now ignoring the presence of normal polarizations, it follows that the susceptibility components that induce those normal polarizations may also be ignored. Therefore, the susceptibility tensors in~\eqref{eq:InvProb} are from now on reduced to $2\times 2$ matrices containing only tangential susceptibility components. 

For convenience and simplicity, the system of equations~\eqref{eq:InvProb} is often cast into a matrix form as
\begin{equation}
\label{eq:InvPropmatrix}
\begin{pmatrix}
\Delta H_y\\
\Delta H_x\\
\Delta E_y\\
\Delta E_x
\end{pmatrix}=
\begin{pmatrix}
\widetilde{\chi}_\text{ee}^{xx} & \widetilde{\chi}_\text{ee}^{xy} & \widetilde{\chi}_\text{em}^{xx} & \widetilde{\chi}_\text{em}^{xy}\\
\widetilde{\chi}_\text{ee}^{yx} & \widetilde{\chi}_\text{ee}^{yy} & \widetilde{\chi}_\text{em}^{yx} & \widetilde{\chi}_\text{em}^{yy}\\
\widetilde{\chi}_\text{me}^{xx} & \widetilde{\chi}_\text{me}^{xy} & \widetilde{\chi}_\text{mm}^{xx} & \widetilde{\chi}_\text{mm}^{xy}\\
\widetilde{\chi}_\text{me}^{yx} & \widetilde{\chi}_\text{me}^{yy} & \widetilde{\chi}_\text{mm}^{yx} & \widetilde{\chi}_\text{mm}^{yy}
\end{pmatrix}
\cdot
\begin{pmatrix}
E_{x,\text{av}}\\
E_{y,\text{av}}\\
H_{x,\text{av}}\\
H_{y,\text{av}}
\end{pmatrix},
\end{equation}
where the tilde susceptibilities have been scaled according to
\begin{subequations}
	\begin{align}
	\te{\chi}_\text{ee} = -j\omega\epsilon_0 \te{\text{N}}\cdot\widetilde{\te{\chi}}_\text{ee}  \quad&\longleftrightarrow\quad \widetilde{\te{\chi}}_\text{ee} = \frac{j}{\omega\epsilon_0} \te{\text{N}}\cdot{\te{\chi}}_\text{ee},\\
	\te{\chi}_\text{mm} = j\omega\mu_0 \te{\text{N}}\cdot\widetilde{\te{\chi}}_\text{mm} \quad&\longleftrightarrow\quad \widetilde{\te{\chi}}_\text{mm} = -\frac{j}{\omega\mu_0} \te{\text{N}}\cdot{\te{\chi}}_\text{mm},\\
	\te{\chi}_\text{em} = -jk_0 \te{\text{N}}\cdot\widetilde{\te{\chi}}_\text{em} \quad&\longleftrightarrow\quad \widetilde{\te{\chi}}_\text{em} = \frac{j}{k_0} \te{\text{N}}\cdot{\te{\chi}}_\text{em},\\
	\te{\chi}_\text{me} = jk_0 \te{\text{N}}\cdot\widetilde{\te{\chi}}_\text{me} \quad&\longleftrightarrow\quad \widetilde{\te{\chi}}_\text{me} = -\frac{j}{k_0} \te{\text{N}}\cdot{\te{\chi}}_\text{me},
	\end{align}
\end{subequations}
with
\begin{equation}
	\te{\text{N}}=
	\begin{pmatrix}
	1 & 0 \\
	0 & -1
	\end{pmatrix}.
\end{equation}

Now that we have established a relationship between the fields interacting with a metasurface and its corresponding susceptibilities with~\eqref{eq:InvPropmatrix}, we shall investigate how the aforementioned fundamental properties generally affect the polarization conversion capabilities of metasurfaces. This may be accomplished most effectively by transforming~\eqref{eq:InvPropmatrix} so that the susceptibilities are related to the metasurface scattering parameters instead of the fields. Indeed, the scattering parameters provide a direct and straightforward connection with polarization effects since they are fully compatible with the Jones calculus formalism~\cite{jones1941new,gupta2015wave}.

Expressing~\eqref{eq:InvPropmatrix} in terms of scattering parameters may be achieved by specifying the incident, reflected and transmitted fields as those of normally propagating plane waves. For instance, a forward propagating (in the +$z$-direction) $x$-polarized incident plane wave may generally be reflected and transmitted as a superposition of both $x$- and $y$-polarized plane waves. The corresponding electric fields are thus given by $\ve{E}_\text{i} = \ve{\hat{x}}$, $\ve{E}_\text{r} = \ve{\hat{x}}\text{S}_{11}^{xx} + \ve{\hat{y}}\text{S}_{11}^{yx}$ and $\ve{E}_\text{t} = \ve{\hat{x}}\text{S}_{21}^{xx} + \ve{\hat{y}}\text{S}_{21}^{yx}$, respectively, where the subscripts 1 and 2 refer to the bottom ($z=0^-$) and top ($z=0^+$) sides of the metasurface. Similar relations may be obtained for a forward propagating $y$-polarized, backward propagating $x$- and $y$-polarized incident plane waves. Substituting these fields, along with their corresponding magnetic counterparts, in~\eqref{eq:InvPropmatrix} and solving for the susceptibilities, yields~\cite{achouri2018design,achouri2020electromagnetic}
\begin{equation}
\label{eq:X}
\begin{split}
&\tilde{\te{\chi}}=
2\begin{pmatrix}
-\frac{\te{\text{N}}}{\eta_0} + \frac{\te{\text{N}}\cdot\te{\text{S}}_{11}}{\eta_0} + \frac{\te{\text{N}}\cdot\te{\text{S}}_{21}}{\eta_0} & -\frac{\te{\text{N}}}{\eta_0} +\frac{\te{\text{N}}\cdot\te{\text{S}}_{12}}{\eta_0} + \frac{\te{\text{N}}\cdot\te{\text{S}}_{22}}{\eta_0} \\
\small{-\te{\text{A}} - \te{\text{A}}\cdot\te{\text{S}}_{11} + \te{\text{A}}\cdot\te{\text{S}}_{21}} & \small{\te{\text{A}} - \te{\text{A}}\cdot\te{\text{S}}_{12}+ \te{\text{A}}\cdot\te{\text{S}}_{22}}
\end{pmatrix}\\
&\qquad\qquad\cdot\begin{pmatrix}
\te{\text{I}} + \te{\text{S}}_{11}+ \te{\text{S}}_{21} &
\te{\text{I}} + \te{\text{S}}_{12}+ \te{\text{S}}_{22}
\\
\frac{\te{\text{J}}}{\eta_0} - \frac{\te{\text{J}}\cdot\te{\text{S}}_{11}}{\eta_0} + \frac{\te{\text{J}}\cdot\te{\text{S}}_{21}}{\eta_0} &
-\frac{\te{\text{J}}}{\eta_0} - \frac{\te{\text{J}}\cdot\te{\text{S}}_{12}}{\eta_0} + \frac{\te{\text{J}}\cdot\te{\text{S}}_{22}}{\eta_0}
\end{pmatrix}^{-1},
\end{split}
\end{equation}
where  $\te{\text{I}}$ is the identity matrix, $\tilde{\te{\chi}}$ is a $4\times 4$ susceptibility matrix corresponding to the one in~\eqref{eq:InvPropmatrix} and
\begin{equation}
\label{eq:Sshape}
\te{\text{S}}_\text{ab}=
\begin{pmatrix}
{\text{S}}_\text{ab}^{xx} & {\text{S}}_\text{ab}^{xy} \\
{\text{S}}_\text{ab}^{yx} & {\text{S}}_\text{ab}^{yy}
\end{pmatrix}, \quad
\te{\text{J}}=
\begin{pmatrix}
0 & -1 \\
1 & 0
\end{pmatrix}, \quad
\te{\text{A}}=
\begin{pmatrix}
0 & 1 \\
1 & 0
\end{pmatrix}.
\end{equation}
Alternatively, it is possible to express the scattering parameters in terms of the susceptibilities as
\begin{equation}
\label{eq:S}
\begin{split}
&\te{\text{S}}= 
\begin{pmatrix}
\frac{\te{\text{N}}}{\eta_0} - \frac{\widetilde{\te{\chi}}_\text{ee}}{2} + \frac{\widetilde{\te{\chi}}_\text{em}\cdot\te{\text{J}}}{2\eta_0} & \frac{\te{\text{N}}}{\eta_0} - \frac{\widetilde{\te{\chi}}_\text{ee}}{2} - \frac{\widetilde{\te{\chi}}_\text{em}\cdot\te{\text{J}}}{2\eta_0} \\
-\te{\text{A}} - \frac{\widetilde{\te{\chi}}_\text{me}}{2} + \frac{\widetilde{\te{\chi}}_\text{mm}\cdot\te{\text{J}}}{2\eta_0} & \te{\text{A}} - \frac{\widetilde{\te{\chi}}_\text{me}}{2} - \frac{\widetilde{\te{\chi}}_\text{mm}\cdot\te{\text{J}}}{2\eta_0}
\end{pmatrix}^{-1}\\
&\qquad\qquad\cdot\begin{pmatrix}
\frac{\widetilde{\te{\chi}}_\text{ee}}{2} + \frac{\te{\text{N}}}{\eta_0}+\frac{\widetilde{\te{\chi}}_\text{em}\cdot\te{\text{J}}}{2 \eta_0} & \frac{\widetilde{\te{\chi}}_\text{ee}}{2} + \frac{\te{\text{N}}}{\eta_0}-\frac{\widetilde{\te{\chi}}_\text{em}\cdot\te{\text{J}}}{2 \eta_0} \\
\frac{\widetilde{\te{\chi}}_\text{me}}{2} + \te{\text{A}}+\frac{\widetilde{\te{\chi}}_\text{mm}\cdot\te{\text{J}}}{2 \eta_0} & \frac{\widetilde{\te{\chi}}_\text{me}}{2} - \te{\text{A}}-\frac{\widetilde{\te{\chi}}_\text{mm}\cdot\te{\text{J}}}{2 \eta_0}
\end{pmatrix}.
\end{split}
\end{equation}
where
\begin{equation}
\te{\text{S}}=
\begin{pmatrix}
\te{\text{S}}_{11} & \te{\text{S}}_{12} \\
\te{\text{S}}_{21} & \te{\text{S}}_{22}
\end{pmatrix}.
\end{equation}

\section{Fundamental Properties of Metasurfaces}
\label{sec:cond}

This section presents the fundamental properties of reciprocity, conservation of energy, rotation invariance and matching and shows how  they may be expressed either in terms of susceptibilities or in terms of scattering parameters. We shall next review and provide the conditions associated with each of these properties. Note that these conditions apply to the very general case of an electromagnetic system and not only to metasurfaces.

\subsection{Reciprocity}

A reciprocal electromagnetic system exhibits the same scattering response when source and receiver are exchanged. From the reciprocity theorem~\cite{Kong2008, calozElectromagneticNonreciprocity2018}, a metasurface is reciprocal if \textit{all} following conditions are satisfied:
\begin{equation}
\label{eq:reciprocity}
\te{\chi}_{\text{ee}} = \te{\chi}_{\text{ee}}^\text{T}, \quad \te{\chi}_{\text{mm}}=\te{\chi}_{\text{mm}}^\text{T}, \quad \te{\chi}_{\text{me}}=-\te{\chi}_{\text{em}}^\text{T},
\end{equation}
where $\text{T}$ is the transpose operation. The corresponding conditions in terms of scattering parameters are~\cite{pozar2011microwave}
\begin{equation}
\label{eq:reciprocityS}
\te{\text{S}}_{11} = \te{\text{S}}_{11}^\text{T}, \quad \te{\text{S}}_{22}=\te{\text{S}}_{22}^\text{T}, \quad \te{\text{S}}_{21}=\te{\text{S}}_{12}^\text{T}.
\end{equation}

Note that it is practically difficult to implement a nonreciprocal electromagnetic system, as it requires the introduction of a time-odd external bias, such as a static magnetic field as the case for Faraday rotators~\cite{ calozElectromagneticNonreciprocity2018}. Therefore, most common electromagnetic systems are de facto reciprocal.

\subsection{Conservation of Energy}

Conservation of energy stipulates that all energy incident on a gainless medium be either scattered or absorbed. If in addition of being gainless, the medium is also lossless\footnote{Although lossless systems do not exist since dissipation is inevitable, ideal design specifications may require losslessness for simplicity, convenience and maximum efficiency.}, then all incident energy must be equal to all the scattered energy. The corresponding conditions in terms of susceptibilities may be deduced from the bianisotropic Poynting theorem as~\cite{lindellMethodsElectromagneticField2000}
\begin{equation}
\label{eq:lossless}
\te{\chi}_{\text{ee}}^*=\te{\chi}_{\text{ee}}^{\text{T}}, \quad \te{\chi}_{\text{mm}}^*=\te{\chi}_{\text{mm}}^{\text{T}}, \quad \te{\chi}_{\text{me}}^*=\te{\chi}_{\text{em}}^{\text{T}}.
\end{equation}
where $*$ is the conjugate operation. The corresponding conditions given in terms of scattering parameters, which have been derived in Appendix~\ref{sec:Sparam},  are given by
\begin{subequations}
	\label{eq:Spow}
	\begin{align}
	&|\text{S}_{12}^{xy}|^2 + |\text{S}_{12}^{yy}|^2 + |\text{S}_{22}^{xy}|^2 + |\text{S}_{22}^{yy}|^2 = 1, \\
	&|\text{S}_{12}^{xx}|^2 + |\text{S}_{12}^{yx}|^2 + |\text{S}_{22}^{xx}|^2 + |\text{S}_{22}^{yx}|^2 = 1, \\
	&|\text{S}_{11}^{xy}|^2 + |\text{S}_{11}^{yy}|^2 + |\text{S}_{21}^{xy}|^2 + |\text{S}_{21}^{yy}|^2 = 1, \\
	&|\text{S}_{11}^{xx}|^2 + |\text{S}_{11}^{yx}|^2 + |\text{S}_{21}^{xx}|^2 + |\text{S}_{21}^{yx}|^2 = 1,
	\end{align}
\end{subequations}
and
\begin{subequations}
	\label{eq:Sarg}
	\begin{align}
	&\text{S}_{11}^{xx*} \text{S}_{11}^{xy} + \text{S}_{11}^{yx*} \text{S}_{11}^{yy} + \text{S}_{21}^{xx*} \text{S}_{21}^{xy} + \text{S}_{21}^{yx*} \text{S}_{21}^{yy} = 0, \\
	&\text{S}_{11}^{xx} \text{S}_{11}^{xy*} + \text{S}_{11}^{yx} \text{S}_{11}^{yy*} + \text{S}_{21}^{xx} \text{S}_{21}^{xy*} + \text{S}_{21}^{yx} \text{S}_{21}^{yy*} = 0, \\
	&\text{S}_{11}^{xx*} \text{S}_{12}^{xx} + \text{S}_{11}^{yx*} \text{S}_{12}^{yx} + \text{S}_{21}^{xx*} \text{S}_{22}^{xx} + \text{S}_{21}^{yx*} \text{S}_{22}^{yx} = 0, \\
	&\text{S}_{11}^{xy*} \text{S}_{12}^{xx} + \text{S}_{11}^{yy*} \text{S}_{12}^{yx} + \text{S}_{21}^{xy*} \text{S}_{22}^{xx} + \text{S}_{21}^{yy*} \text{S}_{22}^{yx} = 0, \\
	&\text{S}_{12}^{xx} \text{S}_{12}^{xy*} + \text{S}_{12}^{yx} \text{S}_{12}^{yy*} + \text{S}_{22}^{xx} \text{S}_{22}^{xy*} + \text{S}_{22}^{yx} \text{S}_{22}^{yy*} = 0, \\
	&\text{S}_{11}^{xx} \text{S}_{12}^{xx*} + \text{S}_{11}^{yx} \text{S}_{12}^{yx*} + \text{S}_{21}^{xx} \text{S}_{22}^{xx*} + \text{S}_{21}^{yx} \text{S}_{22}^{yx*} = 0, \\
	&\text{S}_{11}^{xy} \text{S}_{12}^{xx*} + \text{S}_{11}^{yy} \text{S}_{12}^{yx*} + \text{S}_{21}^{xy} \text{S}_{22}^{xx*} + \text{S}_{21}^{yy} \text{S}_{22}^{yx*} = 0, \\
	&\text{S}_{11}^{xx*} \text{S}_{12}^{xy} + \text{S}_{11}^{yx*} \text{S}_{12}^{yy} + \text{S}_{21}^{xx*} \text{S}_{22}^{xy} + \text{S}_{21}^{yx*} \text{S}_{22}^{yy} = 0, \\
	&\text{S}_{11}^{xy*} \text{S}_{12}^{xy} + \text{S}_{11}^{yy*} \text{S}_{12}^{yy} + \text{S}_{21}^{xy*} \text{S}_{22}^{xy} + \text{S}_{21}^{yy*} \text{S}_{22}^{yy} = 0, \\
	&\text{S}_{12}^{xx*} \text{S}_{12}^{xy} + \text{S}_{12}^{yx*} \text{S}_{12}^{yy} + \text{S}_{22}^{xx*} \text{S}_{22}^{xy} + \text{S}_{22}^{yx*} \text{S}_{22}^{yy} = 0, \\
	&\text{S}_{11}^{xx} \text{S}_{12}^{xy*} + \text{S}_{11}^{yx} \text{S}_{12}^{yy*} + \text{S}_{21}^{xx} \text{S}_{22}^{xy*} + \text{S}_{21}^{yx} \text{S}_{22}^{yy*} = 0, \\
	&\text{S}_{11}^{xy} \text{S}_{12}^{xy*} + \text{S}_{11}^{yy} \text{S}_{12}^{yy*} + \text{S}_{21}^{xy} \text{S}_{22}^{xy*} + \text{S}_{21}^{yy} \text{S}_{22}^{yy*} = 0.
	\end{align}
\end{subequations}
These conditions thus require that not only the amplitude of the scattering parameters must be related to each other but also their phase.

\subsection{Rotation Invariance}

In the context of this work, rotation invariance implies that the scattering response of a system remains identical irrespectively of its angular orientation in a plane transverse to that of wave propagation. 

A metasurface is rotation invariant if all of its susceptibilities and scattering tensors, expressed as $2\times 2$ matrices as in Sec.~\ref{sec:GSTC}, satisfy the condition
\begin{equation}
\label{eq:RotInvar}
\te{\text{M}} = \te{\text{R}}(\phi)\cdot\te{\text{M}}\cdot\te{\text{R}}^\text{T}(\phi) = \begin{pmatrix}
A && B \\
-B && A
\end{pmatrix},
\end{equation}
where $\te{\text{M}}$ represents either a susceptibility or a scattering matrix with $A,B \in \mathbb{C}$ and $\te{\text{R}}(\phi)$ is the rotation matrix defined as
\begin{equation}
\label{eq:rotmat}
\te{\text{R}}(\phi) = \begin{pmatrix}
\cos{\phi} && -\sin{\phi} \\
\sin{\phi} && \cos{\phi}
\end{pmatrix}.
\end{equation}

Note that the direct connection between susceptibilities and scattering parameters provided by relations~\eqref{eq:X} and~\eqref{eq:S} implies that if $\te{\chi}_\text{ee}$, $\te{\chi}_\text{mm}$, $\te{\chi}_\text{me}$ and $\te{\chi}_\text{em}$ all simultaneously satisfy the condition~\eqref{eq:RotInvar}, then $\te{\text{S}}_{11}, \te{\text{S}}_{22}, \te{\text{S}}_{21}$ and $\te{\text{S}}_{12}$ also satisfy it, and vice versa.

\subsection{Matching}

Matching consists in canceling all reflection from a system. Using the formalism developed in Sec.~\ref{sec:GSTC}, a matched metasurface is thus reflectionless, i.e., 
\begin{equation}
\label{eq:Smatch}
	\te{\text{S}}_{11} = 0~\text{and}~\te{\text{S}}_{22} =0,
\end{equation}
which implies that certain conditions in terms of susceptibilities be satisfied. These conditions may be derived by substituting~\eqref{eq:Smatch} into~\eqref{eq:X}, leaving $\te{\text{S}}_{12}$ and $\te{\text{S}}_{21}$ as free parameters, and grouping the remaining terms together to obtain
\begin{subequations}
	\label{eq:reflectionless}
	\begin{align}
	\te{\chi}_\text{mm} &= -\te{\text{J}}\cdot\te{\chi}_\text{ee} \cdot\te{\text{J}},\label{eq:reflectionlessXm}\\
	\te{\chi}_\text{me} &= \te{\text{J}}\cdot\te{\chi}_\text{em}\cdot \te{\text{J}},\label{eq:reflectionless2}
	\end{align}
\end{subequations}
which corresponds to a generalization of the Kerker conditions for a bianisotropic metasurface~\cite{Kerker2013}. Note that if the metasurface is also reciprocal, i.e., if conditions~\eqref{eq:reciprocity} are satisfied, then~\eqref{eq:reflectionless2} reduces to
\begin{equation}
\label{eq:reflectionless3}
\te{\chi}_\text{em} = 	-\te{\chi}_\text{me}^\text{T} = \kappa \te{\text{I}},
\end{equation}
where $\kappa$ is the chirality parameter.

\section{Combined Properties and their Effects on Polarization Conversion}
\label{sec:combined}

\subsection{Susceptibility Conditions for Metasurfaces with Combined Properties}
\label{sec:comb2}
This section explains how the fundamental properties presented in Sec.~\ref{sec:cond} may affect the design of a metasurface, for instance, by restricting the type of electromagnetic transformations that it may accomplish. For this purpose, we shall now derive and investigate the conditions, given in terms of susceptibilities, so that a metasurface simultaneously satisfies one or several of these properties. Note that because the matching property strongly limits the breadth of applications that a metasurface may realize, we will for now restrict our attention to the properties of reciprocity, energy conservation and rotation invariance.

\begin{figure*}[h!]
	\centering
	\includegraphics[width=0.75\textwidth]{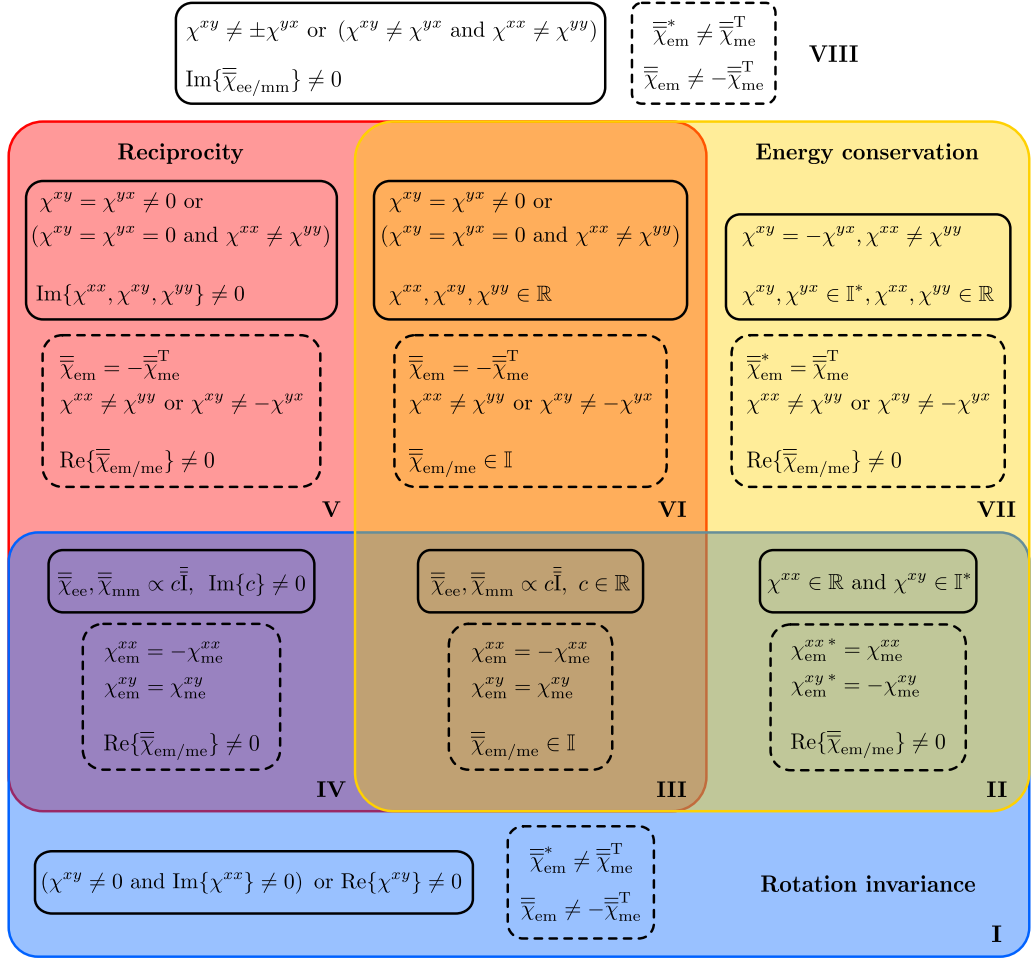}
	\caption{Venn diagram showing different associations of the conditions of reciprocity, energy conservation and rotation invariance. Refer to Sec.~\ref{sec:comb2} for more details.}
	\label{fig:RRL}
\end{figure*}

Using different combinations of relations~\eqref{eq:reciprocity},~\eqref{eq:lossless} and~\eqref{eq:RotInvar}, we obtain the general Venn diagram presented in Fig.~\ref{fig:RRL}.
This diagram consists of 8 different regions, each providing the conditions that a given metasurface must satisfy to exhibit the corresponding properties. It follows that in regions I, V and VII only one condition must be satisfied out of the three considered, e.g., a metasurface is rotation invariant in region I but it is simultaneously nonreciprocal and does not satisfy energy conservation. In regions II, IV and VI two conditions are satisfied, while in region III all conditions must be satisfied. Finally, a metasurface that would be classified in region VIII violates all three conditions.

The classification provided in Fig.~\ref{fig:RRL} assumes that the susceptibility matrices have the following generic form:
\begin{equation}
\label{eq:genMat}
\te{\chi}=
\begin{pmatrix}
\chi^{xx} && \chi^{xy}\\
\chi^{yx} && \chi^{yy}
\end{pmatrix},
\end{equation}
where $\te{\chi}$ is either  $\te{\chi}_\text{ee}$, $\te{\chi}_\text{mm}$, $\te{\chi}_\text{me}$ or $\te{\chi}_\text{em}$. The conditions surrounded by a solid black line apply to $\te{\chi}_\text{ee}$ and $\te{\chi}_\text{mm}$, while those surrounded by a dashed black line apply to $\te{\chi}_\text{me}$ and $\te{\chi}_\text{em}$. \textit{Note that at least one of these two types of conditions must be satisfied for a metasurface to be classified within a given region}. Obviously, if a metasurface does not possess bianisotropic susceptibilities, i.e., $\te{\chi}_\text{em}=\te{\chi}_\text{me}=0$, then the conditions that are surrounded by a dashed black line should be simply ignored. 

We emphasize that for a metasurface to be nonreciprocal, it is sufficient that \textit{at least one} of the conditions in~\eqref{eq:reciprocity} be violated. The same applies to the energy conservation and rotation invariance conditions~\eqref{eq:lossless} and~\eqref{eq:RotInvar}, respectively. This has the following important consequence: assume, for instance, an anisotropic metasurface ($\te{\chi}_\text{em}=\te{\chi}_\text{me}=0$), and consider the condition on $\te{\chi}_\text{ee}$ and $\te{\chi}_\text{mm}$ that $\text{Im}\{c\}\neq 0$ in region IV. The purpose of this condition is to ensure that the metasurface violates conservation of energy, thus classifying it in region IV instead of region III. However, this condition needs not necessarily apply on both $\te{\chi}_\text{ee}$ and $\te{\chi}_\text{mm}$ simultaneously for the metasurface to violate conservation of energy. Indeed, it would be sufficient that either $\te{\chi}_\text{ee}$ or $\te{\chi}_\text{mm}$ contains nonzero imaginary parts for the metasurface to be classified in region IV. Similarly, for a bianisotropic metasurface, assuming the conditions on $\te{\chi}_\text{em}$ and $\te{\chi}_\text{me}$ in region IV are satisfied, then the condition $\text{Im}\{c\}\neq 0$ on $\te{\chi}_\text{ee}$ and $\te{\chi}_\text{mm}$ does not necessarily need to be satisfied since $\text{Re}\{\te{\chi}_\text{em/me}\}\neq 0$ already ensures that the metasurface violates conservation of energy. This specificity generalizes to all $\text{Re}\{\cdot\}\neq 0$ and $\text{Im}\{\cdot\}\neq 0$ conditions in the diagram. On the other hand, the conditions $\cdot \in \mathbb{R}$, $\cdot \in \mathbb{I}$  and $\cdot \in \mathbb{I}^*$, that appear in regiones II, III, VI and VII, must be satisfied.

\subsection{Effect on Polarization Conversion}
\label{sec:effectpol}

Now that we have established how the properties of reciprocity, energy conservation and rotation invariance combine with each other through the conditions in Fig.~\ref{fig:RRL}, we will illustrate how these conditions may affect the scattering response of a metasurface. For this purpose, we next consider a series of metasurface synthesis examples, where, for each region in Fig.~\ref{fig:RRL}, we specify a desired set of scattering parameters and solve~\eqref{eq:X} for the corresponding susceptibilities. To be consistent with the specifications of uniformity and normal incidence imposed to derive~\eqref{eq:X}, we next restrict our attention to synthesis examples corresponding to linear-to-linear and linear-to-circular polarization conversions. For simplicity, we also specify the synthesized metasurfaces to be reflectionless so that relations~\eqref{eq:Smatch} and~\eqref{eq:reflectionless} are satisfied. This reduces the number of unknowns in the synthesis problems since we do not have to specify specific values for the reflected fields and only have to specify the transmitted ones. 

We provide the following examples for each regions in Fig.~\ref{fig:RRL}:

\subsubsection{Example for region I}

\begin{figure}[h!]
	\centering
	\includegraphics[width=0.6\columnwidth]{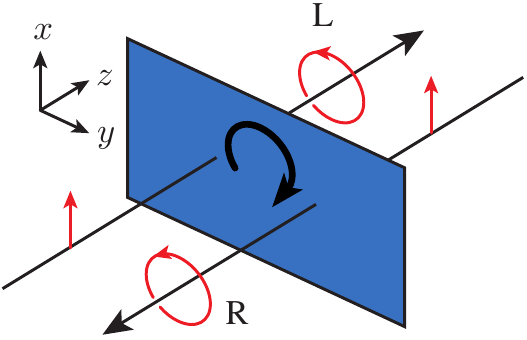}
	\caption{Polarization transformations according to~\eqref{eq:S1}. The black arrow indicates that the metasurface is rotation invariant.}
	\label{fig:pol1}
\end{figure}

The transmission scattering matrices are specified to be
\begin{equation}
\label{eq:S1}
\te{\text{S}}_{21} = 
\frac{\sqrt{2}}{2}
	\begin{pmatrix}
	1 && j \\
	-j && 1
	\end{pmatrix}
	=\te{\text{S}}_{12},
\end{equation}
which corresponds to a linear-to-circular nonreciprocal transformation, as shown in Fig.~\ref{fig:pol1}. 
The corresponding susceptibilities are obtained by substituting~\eqref{eq:S1}, along with $\te{\text{S}}_{11} =\te{\text{S}}_{22} =0$, into~\eqref{eq:X}, which yields
\begin{equation}
\label{eq:X1}
\te{\chi}_\text{ee} = 
\frac{2}{k_0}
\begin{pmatrix}
j(\sqrt{2} - 1) &&\sqrt{2} - 2 \\
2-\sqrt{2}  && j(\sqrt{2}-2)
\end{pmatrix},
\end{equation}
where the magnetic susceptibility tensor is omitted here for convenience but may be computed using~\eqref{eq:reflectionlessXm} and $\te{\chi}_\text{em}=\te{\chi}_\text{me}=0$. It is straightforward to verify that~\eqref{eq:S1} and~\eqref{eq:X1} both satisfy~\eqref{eq:RotInvar} making the metasurface rotation invariant, while also violating both reciprocity~\eqref{eq:reciprocity} and energy conservation~\eqref{eq:lossless}.

\subsubsection{Example for region II}

\begin{figure}[h!]
	\centering
	\includegraphics[width=0.6\columnwidth]{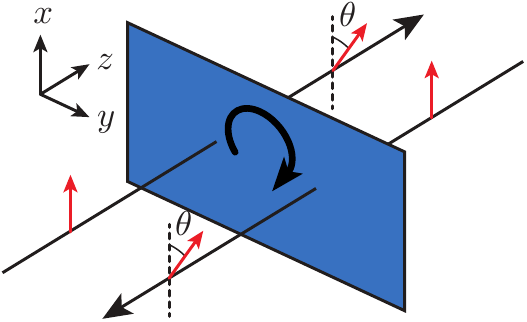}
	\caption{Polarization transformations according to~\eqref{eq:S2}. The black arrow indicates that the metasurface is rotation invariant.}
	\label{fig:pol2}
\end{figure}

The transmission scattering matrices are specified to be
\begin{equation}
\label{eq:S2}
\te{\text{S}}_{21} = 
\begin{pmatrix}
\cos{\theta} && -\sin{\theta} \\
\sin{\theta} && \cos{\theta}
\end{pmatrix}
=\te{\text{S}}_{12},
\end{equation}
which corresponds to a linear-to-linear nonreciprocal rotation of polarization, where $\theta$ is the polarization rotation angle, as shown in Fig.~\ref{fig:pol2}. 
As before, the corresponding susceptibilities are obtained from~\eqref{eq:X}, as
\begin{equation}
\te{\chi}_\text{ee} = 
\frac{2j}{k_0}\tan{\left(\frac{\theta}{2}\right)}
\begin{pmatrix}
0 && -1 \\
1  && 0
\end{pmatrix},
\end{equation}
which indeed satisfies the condition in region II since $A=0 \in \mathbb{R}$ and $B \in \mathbb{I^*}$.

\subsubsection{Example for region III}

\begin{figure}[h!]
	\centering
	\includegraphics[width=0.6\columnwidth]{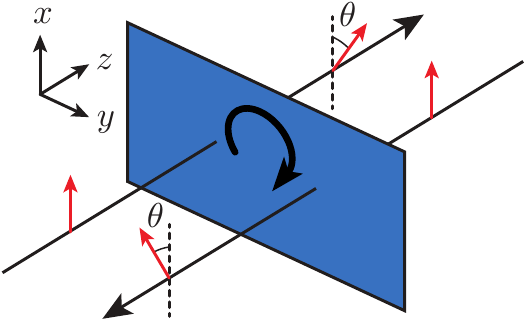}
	\caption{Polarization transformations according to~\eqref{eq:S3}. The black arrow indicates that the metasurface is rotation invariant.}
	\label{fig:pol3}
\end{figure}

The transmission scattering matrices are specified to be
\begin{equation}
\label{eq:S3}
\te{\text{S}}_{21} = 
\begin{pmatrix}
\cos{\theta} && -\sin{\theta} \\
\sin{\theta} && \cos{\theta}
\end{pmatrix}
=\te{\text{S}}_{12}^\text{T},
\end{equation}
which corresponds to a linear-to-linear rotation of polarization, similar to the one given in~\eqref{eq:S2} but that is reciprocal since~\eqref{eq:S3} satisfies~\eqref{eq:reciprocity}. An illustration of its scattering response is shown in Fig.~\ref{fig:pol3}. 
The corresponding susceptibilities are
\begin{equation}
\label{eq:X3}
\te{\chi}_\text{em} = 
\frac{2j}{k_0}\tan{\left(\frac{\theta}{2}\right)}\te{\text{I}},
\end{equation}
where $\te{\chi}_\text{ee}=\te{\chi}_\text{mm}=0$ and $\te{\chi}_\text{me}$ is given by~\eqref{eq:reflectionless3}. 

The susceptibilities~\eqref{eq:X3} correspond to a chiral metasurface~\cite{6477089}, which would rotate the polarization of a linearly polarized incident wave by an angle $\theta$ irrespectively of the angular orientation of the metasurface. Since such a metasurface is reciprocal, reflectionless, lossless, gainless and rotation invariant, it represents the most practical design for a polarization rotation operation.

It is interesting to note that it is impossible to achieve a liner-to-circular polarization conversion with a reciprocal, reflectionless, lossless, gainless and rotation invariant metasurface. To demonstrate this, consider that for a reciprocal, reflectionless and rotation invariant metasurface, the energy conservation conditions in~\eqref{eq:Sarg} and~\eqref{eq:Spow} reduce to
\begin{equation}
\label{eq:AT}
|T_\text{co}|^2 + |T_\text{cross}|^2 = 1,
\end{equation}
and
\begin{equation}
\label{eq:ArT}
T_\text{co}T_\text{cross}^* = T_\text{co}^*T_\text{cross},
\end{equation}
where $T_\text{co} = \text{S}_{21}^{xx} = \text{S}_{21}^{yy}$ is the co-polarized transmission coefficient and $T_\text{cross} = \text{S}_{21}^{xy} = -\text{S}_{21}^{yx}$ is the cross-polarized transmission coefficient. We now directly see that a linear-to-circular polarization conversion, as the one specified by the scattering matrix~\eqref{eq:S1}, would satisfy~\eqref{eq:AT} but would violate~\eqref{eq:ArT}, implying that such a metasurface would require active and/or lossy scattering particles.

\subsubsection{Example for region IV}

\begin{figure}[h!]
	\centering
	\includegraphics[width=0.6\columnwidth]{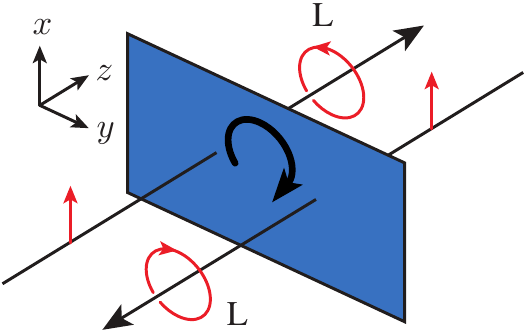}
	\caption{Polarization transformations according to~\eqref{eq:S4}. The black arrow indicates that the metasurface is rotation invariant.}
	\label{fig:pol4}
\end{figure}

We now compute the susceptibilities of a reciprocal, reflectionless and rotation invariant metasurface that performs a linear-to-circular polarization conversion, as shown in Fig.~\ref{fig:pol4}. 
The transmission scattering matrices are specified to be
\begin{equation}
\label{eq:S4}
\te{\text{S}}_{21} = 
\frac{\sqrt{2}}{2}
\begin{pmatrix}
1 && j \\
-j && 1
\end{pmatrix}
=\te{\text{S}}_{12}^\text{T},
\end{equation}
which leads to 
\begin{subequations}
\begin{align}
\te{\chi}_\text{ee} &= 
\frac{2j}{k_0}(\sqrt{2} - 1)\te{\text{I}},\\
\te{\chi}_\text{em} &= 
\frac{2}{k_0}(2-\sqrt{2})\te{\text{I}},
\end{align}
\end{subequations}
which indeed violates conservation of energy, as explained above.

The 4 remaining examples consist of metasurfaces that are rotation dependent meaning that their scattering and susceptibility matrices do not satisfy~\eqref{eq:RotInvar}. 

\subsubsection{Example for region V}

\begin{figure}[h!]
	\centering
	\includegraphics[width=0.6\columnwidth]{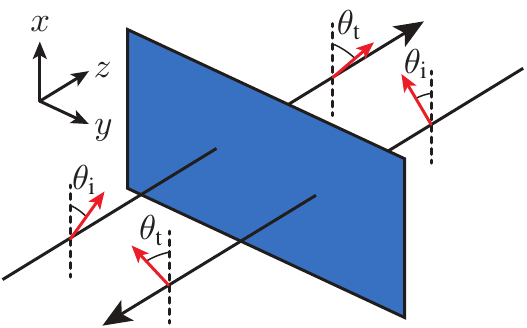}
	\caption{Polarization transformations according to~\eqref{eq:S5}.}
	\label{fig:pol5}
\end{figure}

We shall next consider the case of a reciprocal linear-to-linear polarization conversion. This time the scattering matrix is derived by considering that the electric field of the incident and transmitted waves are $\ve{E}_\text{i} = \ve{\hat{x}}\cos\theta_\text{i} + \ve{\hat{y}}\sin\theta_\text{i}$ and $\ve{E}_\text{t} = \ve{\hat{x}}\cos\theta_\text{t} + \ve{\hat{y}}\sin\theta_\text{t}$, respectively. It follows that the scattering matrix relating these two fields is simply given by
\begin{equation}
\label{eq:S5}
\te{\text{S}}_{21} = 
\begin{pmatrix}
\sec{\theta_\text{i}}\cos{\theta_\text{t}} && 0 \\
0 && \csc{\theta_\text{i}}\sin{\theta_\text{t}}
\end{pmatrix}
=\te{\text{S}}_{12}^\text{T}.
\end{equation}
An illustration of this transformation is depicted in Fig.~\ref{fig:pol5}. 
The corresponding susceptibilities are
\begin{equation}
\te{\chi}_\text{ee} = 
-\frac{2j}{k_0}
\begin{pmatrix}
\frac{\cos{\theta_\text{i}}-\cos{\theta_\text{t}}}{\cos{\theta_\text{i}}+\cos{\theta_\text{t}}} && 0 \\
0  && \frac{\sin{\theta_\text{i}}-\sin{\theta_\text{t}}}{\sin{\theta_\text{i}}+\sin{\theta_\text{t}}}
\end{pmatrix}.
\end{equation}
This clearly shows that the susceptibilities are dependent on the orientation of the fields and that rotation of the metasurface would yield a different scattering response.

\subsubsection{Example for region VI}

\begin{figure}[h!]
	\centering
	\includegraphics[width=0.6\columnwidth]{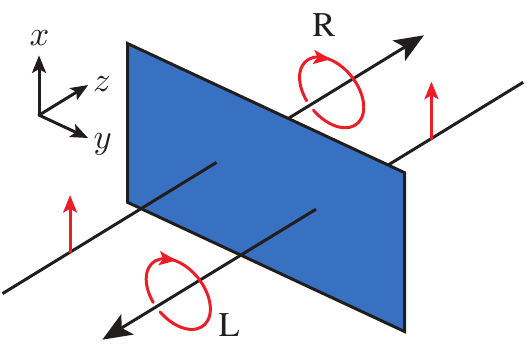}
	\caption{Polarization transformations according to~\eqref{eq:S6}.}
	\label{fig:pol6}
\end{figure}

We now consider the case of a reciprocal quarter-wave plate metasurface oriented so that it transforms an $x$-polarized incident wave into a right-handed circularly polarized transmitted wave, as shown in Fig.~\ref{fig:pol6}.
The corresponding scattering matrix is~\cite{jones1941new}
\begin{equation}
\label{eq:S6}
\te{\text{S}}_{21} = 
\frac{\sqrt{2}}{2}
\begin{pmatrix}
1 && j \\
j && 1
\end{pmatrix}
=\te{\text{S}}_{12}^\text{T},
\end{equation}
and the associated susceptibilities are
\begin{equation}
\label{eq:X06}
\te{\chi}_\text{ee} = 
\frac{2}{k_0}(1 - \sqrt{2})
\begin{pmatrix}
0 && 1 \\
1  && 0
\end{pmatrix}.
\end{equation}
It is interesting to note that, while a quarter-wave plate is a birefringent medium that should be described in terms of a susceptibility matrix with different diagonal components, it is here given in terms of off-diagonal components. This is due to the relative orientation of the metasurface with respect to the incident field, i.e., its fast axis is oriented at $45^\circ$ in the $xy$-plane instead of being aligned along the $x$- or $y$-axis. To demonstrate this, we next rotate the metasurface by $45^\circ$ so that its fast axis is oriented along the $y$-axis. Using the rotation matrix~\eqref{eq:rotmat}, the scattering matrix~\eqref{eq:S6} becomes
\begin{equation}
\te{\text{S}}_{21} = 
e^{j\frac{\pi}{4}}
\begin{pmatrix}
-j && 0 \\
0 && 1
\end{pmatrix}
=\te{\text{S}}_{12}^\text{T},
\end{equation}
and its susceptibilities are
\begin{equation}
\label{eq:X6}
\te{\chi}_\text{ee} = 
\frac{2}{k_0}(1-\sqrt{2})
\begin{pmatrix}
-1 && 0 \\
0  && 1
\end{pmatrix}.
\end{equation}
Now that the slow and fast axes of this quarter-wave plate metasurface are respectively aligned with the $x$- and $y$-axis, the retrieved susceptibility matrix~\eqref{eq:X6} is diagonal, as expected. Note the important difference between~\eqref{eq:X06} and~\eqref{eq:X6}, which is due to the fact that the metasurface is not rotation invariant.

\subsubsection{Example for region VII}

\begin{figure}[h!]
	\centering
	\includegraphics[width=0.6\columnwidth]{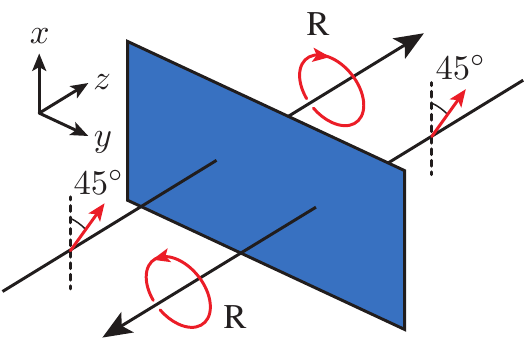}
	\caption{Polarization transformations according to~\eqref{eq:S7} assuming incident waves polarized at $45^\circ$.}
	\label{fig:pol7}
\end{figure}

We again consider the case of a diagonally oriented quarter-wave plate metasurface with a scattering matrix given by
\begin{equation}
\label{eq:S7}
\te{\text{S}}_{21} =
e^{j\frac{\pi}{4}} 
\begin{pmatrix}
1 && 0 \\
0 && j
\end{pmatrix}
=-\te{\text{S}}_{12}^{-1}.
\end{equation}
An illustration of the scattering response of this metasurface is illustrated in Fig.~\ref{fig:pol7} for a obliquely polarized incident plane wave.
Its susceptibilities are given by
\begin{subequations}
	\label{eq:X7}
\begin{align}
\te{\chi}_\text{ee} &= 
-\frac{2\sqrt{2}}{k_0}\te{\text{I}},\\
\te{\chi}_\text{em} &= 
\frac{2}{k_0}
\begin{pmatrix}
0 && 1 \\
1  && 0
\end{pmatrix}.
\end{align}
\end{subequations}
In this example, $\te{\chi}_\text{ee}$ does not satisfy the conditions given in region VII (in black solid line), while $\te{\chi}_\text{em}$ and $\te{\chi}_\text{me}$ do satisfy them. In fact, the susceptibility tensor $\te{\chi}_\text{ee}$, on its own, satisfies the conditions of reciprocity, energy conservation and rotation invariance. It follows that the bianisotropic metasurface given by the susceptibility tensors~\eqref{eq:X7} is classified in region VII only because its $\te{\chi}_\text{em}$ and $\te{\chi}_\text{me}$ tensors violate the conditions of reciprocity and rotation invariance.

\subsubsection{Example for region VIII}

\begin{figure}[h!]
	\centering
	\includegraphics[width=0.6\columnwidth]{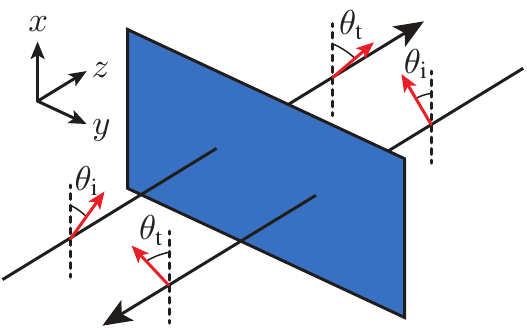}
	\caption{Polarization transformations according to~\eqref{eq:S8}.}
	\label{fig:pol8}
\end{figure}

Finally, we synthesize a nonreciprocal, active and/or lossy and rotation dependent metasurface that rotates the polarization of a linearly polarized incident wave. The corresponding scattering matrices are, from~\eqref{eq:S5}, given by
\begin{equation}
\label{eq:S8}
\te{\text{S}}_{21} = 
\begin{pmatrix}
\sec{\theta_\text{i}}\cos{\theta_\text{t}} && 0 \\
0 && \csc{\theta_\text{i}}\sin{\theta_\text{t}}
\end{pmatrix}
=\te{\text{S}}_{12}^{-1}.
\end{equation}
The corresponding scattering response is shown in Fig.~\ref{fig:pol8} and its susceptibilities are
\begin{equation}
\te{\chi}_\text{ee} = 
\frac{2j}{k_0}
\begin{pmatrix}
0 && \frac{\cos{\theta_\text{t}}-\cos{\theta_\text{i}}}{\cos{\theta_\text{i}}+\cos{\theta_\text{t}}} \\
\frac{\sin{\theta_\text{i}}-\sin{\theta_\text{t}}}{\sin{\theta_\text{i}}+\sin{\theta_\text{t}}}  && 0
\end{pmatrix}.
\end{equation}

\section{Symmetry Breaking Effects on Polarization}
\label{sec:sym}

We have seen in Sec.~\ref{sec:effectpol} several polarization converting metasurface designs, specified in terms of scattering parameters, and how their electromagnetic properties may be classified according to the diagram in Fig.~\ref{fig:RRL}. We shall now investigate how some of these metasurface may be practically implemented in terms of actual scattering particles. For this purpose, we will next restrict our attention to reciprocal and gainless metasurfaces since nonreciprocal and/or active metasurfaces are practically difficult to realize, especially in the optical regime where fabrication technologies are more limiting than those in the microwave regime. This implies that, in practice, most metasurfaces may be classified within regions III to VI in Fig.~\ref{fig:RRL}. Note that in terms of design specifications, a metasurface is generally specified to be lossless\footnote{In addition of being reciprocal and gainless.} to maximize its efficiency even though it is ultimately fabricated with materials that necessarily exhibit loss. It follows that most \textit{ideal} metasurface designs would be classified within regions III or VI, while the corresponding physical structures would be classified within regions IV or V due to their inherent Ohmic or dielectric losses.

In order to design the scattering particles of a metasurface, we must understand how their shape is related to the metasurface effective susceptibilities and, equivalently, to the metasurface scattering response. As we shall next demonstrate, it turns out that the structural symmetries of the scattering particles are directly related to their effects on the state of polarization of the waves that a metasurface scatters.

To illustrate the relationships between scattering particle shape, susceptibilities and scattering response, we next consider the scattering particles proposed in~\cite{menzel_advanced_2010,kruk2020tailoring} from which we identify 7 types of distinct scattering responses that we shall next investigate. Since the connection between the structural symmetries of these scattering particles and their corresponding scattering responses, given in terms of Jones matrices, is already provided in~\cite{menzel_advanced_2010,kruk2020tailoring}, we next limit ourselves to establishing a connection between their provided Jones matrix and their effective susceptibilities. To do so, we assume a simplified scenario for convenience. 

\begin{figure*}[h!]
	\centering
	\includegraphics[width=0.75\textwidth]{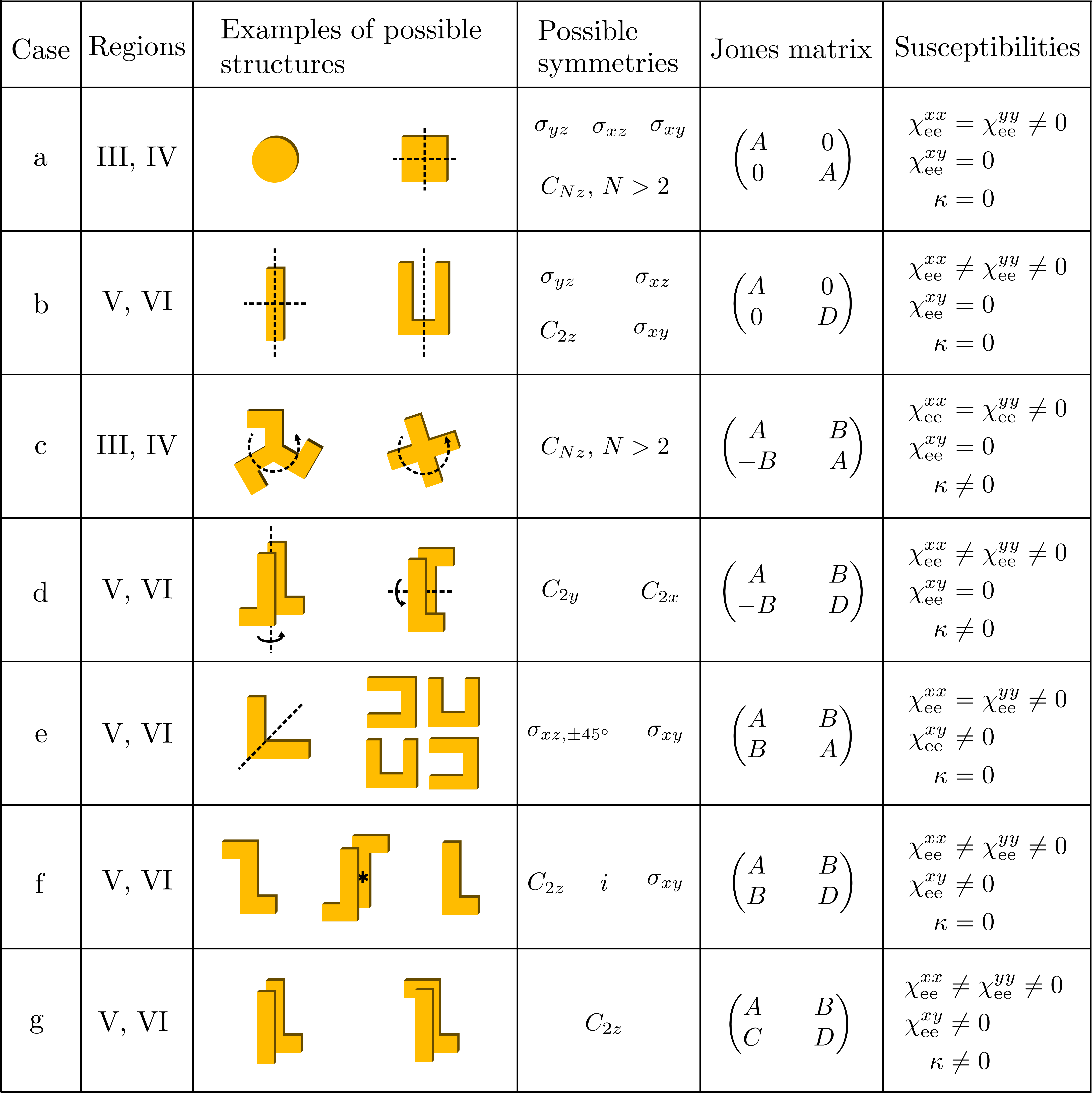}
	\caption{Relationships between scattering particle shapes and their corresponding symmetries, Jones matrix and effective susceptibilities. These apply to a metasurface made of a subwavelength periodic square lattice extending in the $x$ and $y$ directions. For each case, we specify in which regions of Fig.~\ref{fig:RRL} would the corresponding metasurface be classified. Note that the susceptibilities shown here are only indicative and some susceptibility components may be missing, e.g., $z$-oriented susceptibility components are not at all considered. Also note that only the Jones matrix of cases a and c are rotation invariant.}
	\label{fig:PolRotSym}
\end{figure*}

Let us therefore consider a reciprocal reflectionless\footnote{We again consider the case of relfectionless metasurfaces to simplify the forthcoming analysis since relations~\eqref{eq:reflectionless} greatly reduce the number of susceptibility unknowns without affecting the general result provided in Fig.~\ref{fig:PolRotSym}.} bianisotropic gainless and lossless metasurface surrounded by vacuum. Inserting the reflectionless conditions~\eqref{eq:reflectionless} and the gainless and lossless conditions~\eqref{eq:lossless} into~\eqref{eq:S} and solving the resulting system for the components of $\te{\text{S}}_{21}$ yields
\begin{subequations}
	\label{eq:JonesPolRot}
	\begin{align}
	A &= \frac{k^2 \left((\chi_\text{ee}^{xy})^2-\kappa^2-\chi_\text{ee}^{xx} \chi_\text{ee}^{yy}\right)+2 j k
		(\chi_\text{ee}^{xx}-\chi_\text{ee}^{yy})-4}{k^2 \left(\kappa ^2+\chi_\text{ee}^{xx}
		\chi_\text{ee}^{yy}-(\chi_\text{ee}^{xy})^2\right)-2 j k (\chi_\text{ee}^{xx}+\chi_\text{ee}^{yy})-4},\\	
	B &= \frac{4 j k (\chi_\text{ee}^{xy}-\kappa )}{k^2 \left(\kappa^2+\chi_\text{ee}^{xx}
		\chi_\text{ee}^{yy}-(\chi_\text{ee}^{xy})^2\right)-2 j k (\chi_\text{ee}^{xx}+\chi_\text{ee}^{yy})-4},\\	
	C &= \frac{4 j k (\kappa +\chi_\text{ee}^{xy})}{k^2 \left(\kappa^2+\chi_\text{ee}^{xx}
		\chi_\text{ee}^{yy}-(\chi_\text{ee}^{xy})^2\right)-2 j k (\chi_\text{ee}^{xx}+\chi_\text{ee}^{yy})-4},\\
	D &= \frac{k^2 \left((\chi_\text{ee}^{xy})^2-\kappa^2-\chi_\text{ee}^{xx} \chi_\text{ee}^{yy}\right)-2 j k
		(\chi_\text{ee}^{xx}-\chi_\text{ee}^{yy})-4}{k^2 \left(\kappa ^2+\chi_\text{ee}^{xx}
		\chi_\text{ee}^{yy}-(\chi_\text{ee}^{xy})^2\right)-2 j k (\chi_\text{ee}^{xx}+\chi_\text{ee}^{yy})-4},		
	\end{align}
\end{subequations}
where $(A,B,C,D) = (S_{21}^{xx}, S_{21}^{xy}, S_{21}^{yx}, S_{21}^{yy})$ for compatibility with the Jones matrix convention, and $\kappa$ is the chirality parameter from~\eqref{eq:reflectionless3}.  The ABCD-matrix formed by the parameters~\eqref{eq:JonesPolRot} is thus the Jones matrix of the metasurface, whose susceptibilities may be expressed by reversing~\eqref{eq:JonesPolRot}, as
\begin{subequations}
	\label{eq:JonesABCD}
	\begin{align}
	\chi_\text{ee}^{xx} &= \frac{2j}{k_0} \left[\frac{A-BC-1+D(A-1)}{A-BC+1+D(A+1)}\right],\\
	\chi_\text{ee}^{yy} &= -\frac{2j}{k_0} \left[\frac{A+BC+1-D(A+1)}{A-BC+1+D(A+1)}\right],\\
	\chi_\text{ee}^{xy} &= \frac{2j}{k_0} \left[\frac{B+C}{A-BC+1+D(A+1)}\right],\\
	\kappa &= \frac{2j}{k_0} \left[\frac{B-C}{A-BC+1+D(A+1)}\right].
	\end{align}
\end{subequations}
Remember that since relations~\eqref{eq:reflectionless} and~\eqref{eq:reflectionless3} are satisfied, we have that $\chi_\text{mm}^{xx} = \chi_\text{ee}^{yy}$, $\chi_\text{mm}^{yy} = \chi_\text{ee}^{xx}$ and $\chi_\text{mm}^{yx} = -\chi_\text{ee}^{xy}$, and $\kappa= \chi_\text{em}^{xx}= \chi_\text{em}^{yy}= -\chi_\text{me}^{xx}=-\chi_\text{me}^{yy}$.

The selected scattering particles along with their structural symmetries as well as the associated Jones matrix and susceptibilities of the corresponding metasurface are presented in Fig.~\ref{fig:PolRotSym}, where the scattering particles are represented as seen from a top-view above the $xy$-plane.
Note that in order to satisfy the reflectionless conditions~\eqref{eq:reflectionless}, the structures presented in Fig.~\ref{fig:PolRotSym} should be at least bi-layered (although most are not presented as such for convenience) in order to induce both electric and magnetic responses and hence cancel reflection~\cite{Kerker2013,achouri2020electromagnetic}. We now discuss these 7 cases individually.
\subsubsection{Case a}
	 These very simple structures typically present mirror symmetries along both the $xz$- and $yz$-planes ($\sigma_{xz}$ and $\sigma_{yz}$) as well as rotation symmetry along the $z$-axis, $C_{Nz}$ with $N>2$. In terms of susceptibilities, they correspond to isotropic media where $\chi_\text{ee}^{xx} = \chi_\text{ee}^{yy}$. Therefore, the Jones matrix of the corresponding metasurface is diagonal with identical $x$-to-$x$ and $y$-to-$y$ responses. While the disk shaped scattering particle exhibits a rotation invariant scattering behavior, it is not the case of the square shaped one, which can only be rotated by multiples of $45^\circ$ to still yield the same effect. Indeed, rotating it by a different angle would lead to an overall metasurface with a lack of mirror symmetries along the $xz$- and $yz$-planes, thus leading to a more complicated polarization effect. Therefore, as they are represented in the figure, these structures do not affect the polarization state of an $x$- or $y$-polarized incident wave and thus $\chi_\text{ee}^{xy} = \kappa = 0$. 
	
	\subsubsection{Case b}
	Generalization of the structures of \textit{case a}, with different dimensions along $x$ and $y$ implying that $\chi_\text{ee}^{xx} \neq \chi_\text{ee}^{yy}$ (birefringence). They exhibit both or only one of the mirror symmetries along the $xz$- and $yz$-planes ($\sigma_{xz}$ and $\sigma_{yz}$) and a rotation symmetry along the $z$-axis, $C_{Nz}$ with $N\leq2$. Their corresponding metasurface Jones matrix is diagonal with different $x$-to-$x$ and $y$-to-$y$. They do not affect the polarization of $x$- or $y$-polarized waves. However, if the incident wave was diagonally polarized, or, equivalently, if the structure was rotated within its unit cell, they would behave as those of \textit{case f} since
	\begin{equation}
	\label{eq:caseb}
	\begin{split}
	&\te{\text{R}}(\phi)\cdot\begin{pmatrix}
	A && 0 \\
	0 && D		\end{pmatrix}\cdot\te{\text{R}}^\text{T}(\phi)= \\
	&\quad \begin{pmatrix}
	A\cos^2{\phi} + D\sin^2{\phi} && (A-D)\cos{\phi}\sin{\phi} \\
	(A-D)\cos{\phi}\sin{\phi} && D\cos^2{\phi} + A\sin^2{\phi}		\end{pmatrix}=\\
	&\qquad\begin{pmatrix}
	A' && B' \\
	B' && D'	
	\end{pmatrix},
	\end{split}
	\end{equation}	
	which would result in polarization conversion effects. Note that in the special case where $\phi=\pi/4$, Eq.~\eqref{eq:caseb} reduces to
	\begin{equation}
	\label{eq:caseb2}
	\begin{split}		
	&\te{\text{R}}(\pi/4)\cdot\begin{pmatrix}
	A && 0 \\
	0 && D		\end{pmatrix}\cdot\te{\text{R}}^\text{T}(\pi/4)=\\
	&\quad \frac{1}{2}\begin{pmatrix}
	A + D && A-D \\
	A-D && A+D		\end{pmatrix}
	=
	\begin{pmatrix}
	A'' && B'' \\
	B'' && A''	
	\end{pmatrix},
	\end{split}
	\end{equation}	
	which corresponds to the response of the structures of \textit{case e}.
	
	\subsubsection{Case c}
\begin{figure}[h!]
		\centering
		\includegraphics[width=0.75\columnwidth]{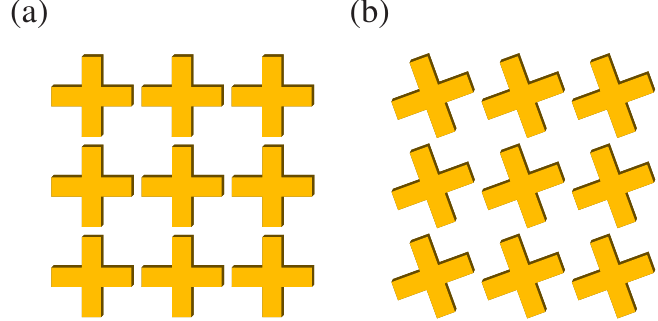}
		\caption{Metasurface made of a periodic arrangement of a cross-shaped structure. (a)~The arms of the crosses are aligned with the main axes of the lattice leading to isotropic scattering response like those of \textit{case a}. (b)~The arms of the crosses are rotated with respect to the axes of the lattice leading to a potential chiral response.}
		\label{fig:chiral}
	\end{figure}

	 Structures exhibiting rotation symmetry so that $C_{Nz}$ with $N>2$ may be used to create chiral media. The first structure has a $C_{3z}$ rotation symmetry and exhibits no $\sigma_{xz}$ or $\sigma_{yz}$ mirror symmetry, while the second one has a $C_{4z}$ rotation symmetry as well as $\sigma_{xz}$ and $\sigma_{yz}$. On their own, these structure are not fundamentally chiral~\cite{calozElectromagneticChiralityPart2020}. To create a chiral medium out of the first structure, it is enough to place it on top of a substrate. This would break the symmetry of the system in the longitudinal direction resulting in an overall chiral response~\cite{menzel_advanced_2010,plum2007giant,calozElectromagneticChiralityPart2020}. 
	 That strategy would not be sufficient for the second cross-shaped structure due to its additional mirror symmetries. However, it is still possible to create a chiral metasurface out of it by placing it on a substrate and rotating the cross within its unit cell. It should be rotated such that its arms are not aligned along the $x$ and $y$ axes or at $\pm45^\circ$ from them. Considering the square lattice of the metasurface being on a $xy$ grid, this rotation of the particle within its cell would effectively cancel the overall $\sigma_{xz}$ and $\sigma_{yz}$ mirror symmetries of the metasurface, hence making it chiral. These two cases are illustrated in Fig.~\ref{fig:chiral}.

	Breaking the longitudinal symmetry of the system may also be achieved by cascading several of these structures and changing their dimensions, composition and orientation. Note that this strategy has also been used to create a stronger chiral response~\cite{plum2007giant}.
	
	This type of chiral medium exhibit a chiral parameter $\kappa \neq 0$, while $\chi_\text{ee}^{xy} = 0$. These are the best type of structure for polarization rotation since their effect is rotation invariant and would accordingly be classified in region III in Fig.~\ref{fig:RRL}.
	
	\subsubsection{Case d}
	 These structures, which possess a $C_{2x}$ or $C_{2y}$ rotation symmetry, result in a generalized chiral response. They induce a counter-rotating effect on the fields, i.e., the metasurface Jones matrix is the negative of its own transpose, like the chiral structures of \textit{case c} but have different diagonal components ($\chi_\text{ee}^{xx} \neq \chi_\text{ee}^{yy}$) due to their different lengths in the $x$- and $y$-directions. Because of that, their effect on the polarization state of the scattered waves depends on the angular orientation of the metasurface.

	\subsubsection{Case e}
	 Structures that have a $\pm 45^\circ$ mirror symmetry with respect to the $xz$- or $yz$-plane exhibit a nonzero $x$-to-$y$ and $y$-to-$x$ coupling leading to polarization conversion but not chirality. As explained in \textit{case b}, such response can be obtained by rotating a birefringent scatterer within its unit cell by a $\pm 45^\circ$ angle. Note that such structures also typically exhibit a mirror symmetry along the $xy$-plane, like the structures of \textit{case a} and \textit{case b}. 
	
	The unit cell composed of 4 split-ring resonators does not, on its own, exhibit a $\pm 45^\circ$ mirror symmetry with respect to the $xz$- or $yz$-plane. However, when considering a metasurface composed of a periodic repetition of this unit cell, we can show that by reflecting the structure diagonally and shifting it by half a period along $x$ or $y$, we retrieve the original unit cell~\cite{menzel_advanced_2010,Decker2009}. Implying that this structure still exhibit the same type of response as the L-shaped one.

	\subsubsection{Case f}
	 A generalization of the structures of \textit{case e} with different $x$-to-$x$ and $y$-to-$y$ responses due to their different lengths in the $x$- and $y$-directions. They can either exhibit a mirror symmetry with respect to the $xy$-plane, like the first and third depicted structures, or an inversion symmetry ($i$), like the second one. The same type of response can also be achieved by rotating a birefringent structure by a given angle, as explained in \textit{case b}.

	\subsubsection{Case g}
	 These structures present either no symmetry ($C_1$) or eventually a $C_{2z}$ rotation symmetry. They can be used to perform any operation on the wave polarization state providing that it does not violate the imposed conditions of reciprocity and energy conservation.

\section{Conclusion}
\label{sec:concl}

This works has presented the general electromagnetic properties of reciprocity, energy conservation, rotation invariance and matching, and provided the associated conditions in terms of susceptibilities and scattering parameters. It has then established how metasurfaces may be classified according to the various possible combinations of these conditions and how this may affect the polarization converting capabilities of metasurfaces. Finally, it has connected the structural symmetries of scattering particles to the corresponding metasurface Jones matrix and susceptibilities.

\section*{Acknowledgements}

We gratefully acknowledge funding from the European Research Council (ERC-2015-AdG-695206 Nanofactory).

\appendices

\section{S-Parameter Conditions for Conservation of Energy}
\label{sec:Sparam}

Consider a uniform gainless and lossless slab\footnote{In the context of this paper, it could be a metasurface.} lying in the $xy$-plane and surrounded by the same medium on both sides. It is simultaneously illuminated by a normally incident plane wave propagating in the $+z$-direction and one propagating in the $-z$-direction, whose electric fields are $\ve{E}_1^{(+)}$ and $\ve{E}_2^{(-)}$, respectively. The bottom and top sides of the slab are denoted with subscripts 1 and 2, while propagation in the $\pm z$-direction is denoted with the superscripts $(\pm)$, respectively. Following the same convention, the fields reflected and transmitted by the slab are
\begin{subequations}
	\label{eq:scatsys}
	\begin{align}
	\begin{pmatrix}
	E_{1,x}^{(-)} \\
	E_{1,y}^{(-)}
	\end{pmatrix}
	&=
	\te{\text{S}}_{11}\cdot	
	\begin{pmatrix}
	E_{1,x}^{(+)} \\
	E_{1,y}^{(+)}
	\end{pmatrix}
	+
	\te{\text{S}}_{12}\cdot	
	\begin{pmatrix}
	E_{2,x}^{(-)} \\
	E_{2,y}^{(-)}
	\end{pmatrix},\\
		\begin{pmatrix}
	E_{2,x}^{(+)} \\
	E_{2,y}^{(+)}
	\end{pmatrix}
	&=
	\te{\text{S}}_{21}\cdot	
	\begin{pmatrix}
	E_{1,x}^{(+)} \\
	E_{1,y}^{(+)}
	\end{pmatrix}
	+
	\te{\text{S}}_{22}\cdot	
	\begin{pmatrix}
	E_{2,x}^{(-)} \\
	E_{2,y}^{(-)}
	\end{pmatrix},
	\end{align}
\end{subequations}
where the scattering matrices have the same form as in~\eqref{eq:Sshape}. 

Since the slab is gainless and lossless, all incident energy must be equal to all scattered energy, which may be expressed as
\begin{equation}
\label{eq:PowCons}
\begin{split}
	|E_{1,x}^{(+)}|^2 &+ |E_{1,y}^{(+)}|^2 + 	|E_{2,x}^{(-)}|^2 + |E_{2,y}^{(-)}|^2 = \\
	 &	|E_{1,x}^{(-)}|^2 + |E_{1,y}^{(-)}|^2 + 	|E_{2,x}^{(+)}|^2 + |E_{2,y}^{(+)}|^2,
\end{split}
\end{equation}
where the terms on the left-hand side are related to the incident energy, while those on the right-hand side are related to the scattered energy. Substituting~\eqref{eq:scatsys} into~\eqref{eq:PowCons} leads to an equation that must be satisfied for any field values. It follows that, by grouping similar terms together, several conditions on the scattering parameters may be derived leading to a total of 4 relations given in terms of the scattering parameters magnitude, provided in~\eqref{eq:Spow}, and 12 relations in terms of their complex values given in~\eqref{eq:Sarg}.

\bibliographystyle{myIEEEtran}
\bibliography{NewLib}

\begin{thebibliography}{10}
\providecommand{\url}[1]{#1}
\csname url@samestyle\endcsname
\providecommand{\newblock}{\relax}
\providecommand{\bibinfo}[2]{#2}
\providecommand{\BIBentrySTDinterwordspacing}{\spaceskip=0pt\relax}
\providecommand{\BIBentryALTinterwordstretchfactor}{4}
\providecommand{\BIBentryALTinterwordspacing}{\spaceskip=\fontdimen2\font plus
\BIBentryALTinterwordstretchfactor\fontdimen3\font minus
  \fontdimen4\font\relax}
\providecommand{\BIBforeignlanguage}[2]{{%
\expandafter\ifx\csname l@#1\endcsname\relax
\typeout{** WARNING: IEEEtran.bst: No hyphenation pattern has been}%
\typeout{** loaded for the language `#1'. Using the pattern for}%
\typeout{** the default language instead.}%
\else
\language=\csname l@#1\endcsname
\fi
#2}}
\providecommand{\BIBdecl}{\relax}
\BIBdecl

\bibitem{achouri2018design}
K.~Achouri and C.~Caloz, ``Design, concepts, and applications of
  electromagnetic metasurfaces,'' \emph{Nanophotonics}, vol.~7, no.~6, pp.
  1095--1116, 2018.

\bibitem{chenHuygensMetasurfacesMicrowaves2018}
M.~Chen, M.~Kim, A.~M. Wong, and G.~V. Eleftheriades, ``Huygens' metasurfaces
  from microwaves to optics: A review,'' \emph{Nanophotonics}, vol.~7, no.~6,
  pp. 1207--1231, Jun. 2018.

\bibitem{asadchy2018bianisotropic}
V.~S. Asadchy, A.~D{\'\i}az-Rubio, and S.~A. Tretyakov, ``Bianisotropic
  metasurfaces: physics and applications,'' \emph{Nanophotonics}, vol.~7,
  no.~6, pp. 1069--1094, 2018.

\bibitem{Sihvola1994}
A.~Sihvola, A.~Viitanen, I.~Lindell, and S.~Tretyakov, \emph{Electromagnetic
  waves in chiral and bi-isotropic media}, ser. The Artech House Antenna
  Library.\hskip 1em plus 0.5em minus 0.4em\relax Artech House, 1994.

\bibitem{serdkov2001electromagnetics}
A.~Serd\t{iu}kov, I.~Semchenko, S.~Tretyakov, and A.~Sihvola,
  \emph{Electromagnetics of bi-anisotropic materials-Theory and
  Application}.\hskip 1em plus 0.5em minus 0.4em\relax Gordon and Breach
  science publishers, 2001, vol.~11.

\bibitem{Asadchy2016a}
V.~S. Asadchy, M.~Albooyeh, S.~N. Tcvetkova, A.~D\'{\i}az-Rubio, Y.~Ra'di, and
  S.~A. Tretyakov, ``Perfect control of reflection and refraction using
  spatially dispersive metasurfaces,'' \emph{Phys. Rev. B}, vol.~94, p. 075142,
  Aug 2016.

\bibitem{8259235}
G.~Lavigne, K.~Achouri, V.~S. Asadchy, S.~A. Tretyakov, and C.~Caloz,
  ``Susceptibility derivation and experimental demonstration of refracting
  metasurfaces without spurious diffraction,'' \emph{IEEE Transactions on
  Antennas and Propagation}, vol.~66, no.~3, pp. 1321--1330, March 2018.

\bibitem{phillion2011lenses}
R.~H. Phillion and M.~Okoniewski, ``Lenses for circular polarization using
  planar arrays of rotated passive elements,'' \emph{IEEE Transactions on
  Antennas and Propagation}, vol.~59, no.~4, pp. 1217--1227, 2011.

\bibitem{lin2013polarization}
J.~Lin, J.~B. Mueller, Q.~Wang, G.~Yuan, N.~Antoniou, X.-C. Yuan, and
  F.~Capasso, ``Polarization-controlled tunable directional coupling of surface
  plasmon polaritons,'' \emph{Science}, vol. 340, no. 6130, pp. 331--334, 2013.

\bibitem{li2014ultra}
M.~Li, L.~Guo, J.~Dong, and H.~Yang, ``An ultra-thin chiral metamaterial
  absorber with high selectivity for lcp and rcp waves,'' \emph{Journal of
  Physics D: Applied Physics}, vol.~47, no.~18, p. 185102, 2014.

\bibitem{kim2016highly}
M.~Kim and G.~V. Eleftheriades, ``Highly efficient all-dielectric optical
  tensor impedance metasurfaces for chiral polarization control,'' \emph{Optics
  letters}, vol.~41, no.~20, pp. 4831--4834, 2016.

\bibitem{arbabi2018full}
E.~Arbabi, S.~M. Kamali, A.~Arbabi, and A.~Faraon, ``Full-stokes imaging
  polarimetry using dielectric metasurfaces,'' \emph{ACS Photonics}, vol.~5,
  no.~8, pp. 3132--3140, 2018.

\bibitem{wu2019tunable}
Z.~Wu, Y.~Ra’di, and A.~Grbic, ``Tunable metasurfaces: A polarization rotator
  design,'' \emph{Physical Review X}, vol.~9, no.~1, p. 011036, 2019.

\bibitem{6477089}
T.~Niemi, A.~Karilainen, and S.~Tretyakov, ``Synthesis of polarization
  transformers,'' \emph{IEEE Trans. Antennas Propag.}, vol.~61, no.~6, pp.
  3102--3111, June 2013.

\bibitem{PhysRevApplied.2.044011}
C.~Pfeiffer and A.~Grbic, ``Bianisotropic metasurfaces for optimal polarization
  control: Analysis and synthesis,'' \emph{Phys. Rev. Applied}, vol.~2, p.
  044011, Oct 2014.

\bibitem{selvanayagam2014polarization}
M.~Selvanayagam and G.~V. Eleftheriades, ``Polarization control using tensor
  huygens surfaces,'' \emph{IEEE Transactions on Antennas and Propagation},
  vol.~62, no.~12, pp. 6155--6168, 2014.

\bibitem{selvanayagam2015design}
M.~Selvanayagam and G.~V. Eleftheriades, ``Design and measurement of tensor
  impedance transmitarrays for chiral polarization control,'' \emph{IEEE
  Transactions on Microwave Theory and Techniques}, vol.~64, no.~2, pp.
  414--428, 2015.

\bibitem{kamali2018review}
S.~M. Kamali, E.~Arbabi, A.~Arbabi, and A.~Faraon, ``A review of dielectric
  optical metasurfaces for wavefront control,'' \emph{Nanophotonics}, vol.~7,
  no.~6, pp. 1041--1068, 2018.

\bibitem{menzel_advanced_2010}
C.~Menzel, C.~Rockstuhl, and F.~Lederer, ``Advanced {{Jones}} calculus for the
  classification of periodic metamaterials,'' \emph{Physical Review A},
  vol.~82, no.~5, p. 053811, Nov. 2010.

\bibitem{kenanakis2012flexible}
G.~Kenanakis, R.~Zhao, A.~Stavrinidis, G.~Konstantinidis, N.~Katsarakis,
  M.~Kafesaki, C.~Soukoulis, and E.~Economou, ``Flexible chiral metamaterials
  in the terahertz regime: a comparative study of various designs,''
  \emph{Optical Materials Express}, vol.~2, no.~12, pp. 1702--1712, 2012.

\bibitem{kruk2020tailoring}
S.~Kruk and Y.~Kivshar, ``Tailoring transmission and reflection with
  metasurfaces,'' in \emph{Dielectric Metamaterials}.\hskip 1em plus 0.5em
  minus 0.4em\relax Elsevier, 2020, pp. 145--174.

\bibitem{achouri2014general}
K.~Achouri, M.~A. Salem, and C.~Caloz, ``General metasurface synthesis based on
  susceptibility tensors,'' \emph{IEEE Trans. Antennas Propag.}, vol.~63,
  no.~7, pp. 2977--2991, Jul. 2015.

\bibitem{angularAchouri2020}
K.~{Achouri} and O.~J.~F. {Martin}, ``Angular scattering properties of
  metasurfaces,'' \emph{IEEE Transactions on Antennas and Propagation},
  vol.~68, no.~1, pp. 432--442, 2020.

\bibitem{achouri2020electromagnetic}
K.~Achouri and C.~Caloz, \emph{Electromagnetic Metasurfaces: Theory and
  Applications}.\hskip 1em plus 0.5em minus 0.4em\relax Wiley-IEEE Press, 2020.

\bibitem{Idemen1973}
M.~M. Idemen, \emph{Discontinuities in the Electromagnetic Field}.\hskip 1em
  plus 0.5em minus 0.4em\relax John Wiley \& Sons, 2011.

\bibitem{kuester2003av}
E.~F. Kuester, M.~Mohamed, M.~Piket-May, and C.~Holloway, ``Averaged transition
  conditions for electromagnetic fields at a metafilm,'' \emph{IEEE Trans.
  Antennas Propag.}, vol.~51, no.~10, pp. 2641--2651, Oct 2003.

\bibitem{Kong2008}
J.~A. Kong, \emph{Electromagnetic Wave Theory}.\hskip 1em plus 0.5em minus
  0.4em\relax EMW Publishing, 2008.

\bibitem{vahabzadehComputationalAnalysisMetasurfaces2018}
Y.~Vahabzadeh, N.~Chamanara, K.~Achouri, and C.~Caloz, ``Computational
  {{Analysis}} of {{Metasurfaces}},'' \emph{IEEE Journal on Multiscale and
  Multiphysics Computational Techniques}, vol.~3, pp. 37--49, 2018.

\bibitem{jones1941new}
R.~C. Jones, ``A new calculus for the treatment of optical systemsi.
  description and discussion of the calculus,'' \emph{J. Opt. Soc. Am.},
  vol.~31, no.~7, pp. 488--493, 1941.

\bibitem{gupta2015wave}
S.~D. Gupta, N.~Ghosh, and A.~Banerjee, \emph{Wave optics: Basic concepts and
  contemporary trends}.\hskip 1em plus 0.5em minus 0.4em\relax CRC Press, 2015.

\bibitem{calozElectromagneticNonreciprocity2018}
C.~Caloz, A.~Al{\`u}, S.~Tretyakov, D.~Sounas, K.~Achouri, and Z.-L.
  {Deck-L{\'e}ger}, ``Electromagnetic {{Nonreciprocity}},'' \emph{Physical
  Review Applied}, vol.~10, no.~4, Oct. 2018.

\bibitem{pozar2011microwave}
D.~Pozar, \emph{Microwave Engineering, 4th Edition}.\hskip 1em plus 0.5em minus
  0.4em\relax Wiley, 2011.

\bibitem{lindellMethodsElectromagneticField2000}
I.~V. Lindell, \emph{Methods for Electromagnetic Field Analysis}, ser. {{IEEE
  Press}} Series on Electromagnetic Wave Theory.\hskip 1em plus 0.5em minus
  0.4em\relax {New York}: {IEEE Press}, 2000.

\bibitem{Kerker2013}
M.~Kerker, \emph{The scattering of light and other electromagnetic radiation:
  physical chemistry: a series of monographs}.\hskip 1em plus 0.5em minus
  0.4em\relax Academic press, 2013, vol.~16.

\bibitem{calozElectromagneticChiralityPart2020}
C.~Caloz and A.~Sihvola, ``Electromagnetic {{Chirality}}, {{Part}} 1: {{The
  Microscopic Perspective}} [{{Electromagnetic Perspectives}}],'' \emph{IEEE
  Antennas and Propagation Magazine}, vol.~62, no.~1, pp. 58--71, Feb. 2020.

\bibitem{plum2007giant}
E.~Plum, V.~Fedotov, A.~Schwanecke, N.~Zheludev, and Y.~Chen, ``Giant optical
  gyrotropy due to electromagnetic coupling,'' \emph{Applied Physics Letters},
  vol.~90, no.~22, p. 223113, 2007.

\bibitem{Decker2009}
M.~Decker, S.~Linden, and M.~Wegener, ``Coupling effects in low-symmetry planar
  split-ring resonator arrays,'' \emph{Opt. Lett.}, vol.~34, no.~10, pp.
  1579--1581, May 2009.

\end{thebibliography}

\end{document}